\documentclass[11pt,preprint]{aastex}
\usepackage{amsmath}
\begin{document}
\title{Probing the cross-power of unresolved cosmic infrared and X-ray backgrounds with upcoming space missions}

\author{A. Kashlinsky\altaffilmark{1}, R. G. Arendt\altaffilmark{2}, N. Cappelluti\altaffilmark{3}, A. Finoguenov\altaffilmark{4}, G. Hasinger\altaffilmark{5}, K. Helgason\altaffilmark{6}, A. Merloni\altaffilmark{7}}
\altaffiltext{1}{
Code 665, Observational Cosmology Lab, NASA Goddard Space Flight Center, 
Greenbelt, MD 20771 and
SSAI, Lanham, MD 20770; email: Alexander.Kashlinsky@nasa.gov} 
\altaffiltext{2}{
Code 665, Observational Cosmology Lab, NASA Goddard Space Flight Center, 
Greenbelt, MD 20771 and CRESST II / University of Maryland Baltimore County, Baltimore, MD 21250}
\altaffiltext{3}{Department of Physics, University of Miami, Coral Gables, Florida 33124}
\altaffiltext{4}{MPE, Giessenbachstr. 1, Garching 85748 Germany and
Department of Physics, University of Helsinki, Helsinki 00014, Finland
}
\altaffiltext{5}{European Space Astronomy Centre, E-28691 Villanueva de la Ca–ada, Madrid, Spain}
\altaffiltext{6}{Science Institute, University of Iceland, IS-107 Reykjavik, Iceland}
\altaffiltext{7}{Max-Planck Institute fuer Extraterrestrische Physik (MPE)
P.O Box 1312 Giessenbachstr. 1,
85741 Garching
Germany }


\def\plotone#1{\centering \leavevmode
\epsfxsize=\columnwidth \epsfbox{#1}}

\def\wisk#1{\ifmmode{#1}\else{$#1$}\fi}

\def\wm2sr {Wm$^{-2}$sr$^{-1}$ }		
\def\nw2m4sr2 {nW$^2$m$^{-4}$sr$^{-2}$\ }		
\def\nwm2sr {nWm$^{-2}$sr$^{-1}$\ }		
\def\nw2m4sr {nW$^2$m$^{-4}$sr$^{-1}$\ }
\def\Ncut {$N_{\rm cut}$\ }
\def\lt     {\wisk{<}}
\def\gt     {\wisk{>}}
\def\le     {\wisk{_<\atop^=}}
\def\ge     {\wisk{_>\atop^=}}
\def\lsim   {\wisk{_<\atop^{\sim}}}
\def\gsim   {\wisk{_>\atop^{\sim}}}
\def\kms    {\wisk{{\rm ~km~s^{-1}}}}
\def\Lsun   {\wisk{{\rm L_\odot}}}
\def\Msun   {\wisk{{\rm M_\odot}}}
\def\um     { $\mu$m\ }
\def\sig    {\wisk{\sigma}}
\def\etal   {{\sl et~al.\ }}
\def\eg	    {{\it e.g.\ }}
\def\ie     {{\it i.e.\ }}
\def\bsl    {\wisk{\backslash}}
\def\by     {\wisk{\times}}
\def\cosec {\wisk{\rm cosec}}
\def\mic {\wisk{ \mu{\rm m }}}

\def\amin   {\wisk{^\prime\ }}
\def\asec   {\wisk{^{\prime\prime}\ }}
\def\cc     {\wisk{{\rm cm^{-3}\ }}}
\def\deg     {\wisk{^\circ}}
\def\ddeg   {\wisk{{\rlap.}^\circ}}
\def\damin  {\wisk{{\rlap.}^\prime}}
\def\dasec  {\wisk{{\rlap.}^{\prime\prime}}}
\def\approxeq{$\sim \over =$}
\def\abouteq{$\sim \over -$}
\def\percm{cm$^{-1}$}
\def\percmsq{cm$^{-2}$}
\def\percmcub{cm$^{-3}$}
\def\perhz{Hz$^{-1}$}
\def\perpc{$\rm pc^{-1}$}
\def\persec{s$^{-1}$}
\def\peryr{yr$^{-1}$}
\def\te{$\rm T_e$}
\def\tenup#1{10$^{#1}$}
\def\to{\wisk{\rightarrow}}
\def\thin{\thinspace}
\def\uk{$\rm \mu K$}
\def\p{\vskip 13pt}

\begin{abstract}
The source-subtracted cosmic infrared background (CIB) fluctuations uncovered in deep {\it Spitzer} data cannot be explained by known galaxy populations and appear strongly coherent with unresolved cosmic X-ray background (CXB). This suggests that the source-subtracted CIB contains emissions from significantly abundant accreting black holes (BHs). We show that theoretically such populations would have the angular power spectrum which is largely independent of the epochs occupied by these sources, provided they are at $z\gsim 4$, offering an important test of the origin of the new populations. Using the current measurements we reconstruct the underlying soft X-ray CXB from the new sources and show that its fluctuations, while consistent with a high-$z$ origin, have an amplitude that cannot be reached in direct measurements with the foreseeable X-ray space missions. This necessitates application of the methods developed by the authors to future IR and X-ray datasets, which must cover large areas of the sky in order to measure the signal with high precision.
The LIBRAE project within ESA's {\it Euclid} mission will probe source-subtracted CIB over $\sim1/2$ the sky at three near-IR bands, and its cross-power with unresolved CXB can be measured then from the concurrent eROSITA mission covering the same areas of the sky. We discuss the required methodology for this measurement and evaluate its projected $S/N$ to show the unique potential of this experimental configuration to accurately probe the CXB from the new BH sources and help identify their epochs.
\end{abstract}

\section{Introduction}
\label{sec:introduction}
The near-IR source-subtracted CIB fluctuations can probe emissions from early stars and black holes (BHs), inaccessible to direct telescopic studies \citep{Kashlinsky:2004,Cooray:2004,Kashlinsky:2005}. Analysis of deep {\it Spitzer} images, specifically assembled for this \citep{Arendt:2010}, revealed source-subtracted CIB fluctuations at 3.6 and 4.5 \mic\ \citep{Kashlinsky:2005a} significantly exceeding those from remaining known galaxy populations \citep{Kashlinsky:2005a,Helgason:2012a} and indicating new cosmological sources. Follow-up studies identified the CIB fluctuation excess to $\sim 1^\circ$ with similar levels across the sky \citep{Kashlinsky:2007a,Kashlinsky:2012,Cooray:2012}. The source-subtracted CIB fluctuations at 3.6 and 4.5 \mic\ appear coherent with soft ([0.5-2]keV) unresolved cosmic X-ray background (CXB) \citep{Cappelluti:2013,Mitchell-Wynne:2016,Cappelluti:2017,Li:2018}. 
The coherence levels indicate a  much larger proportion of accreting BHs among the new sources than in known populations \citep{Helgason:2014}. Two suggestions have been made for the origin of these populations, both at high $z$: 1) direct collapse BHs \citep[][]{Yue:2013} and 2) LIGO-type primordial BHs making up dark matter \citep{Kashlinsky:2016}. See review by \cite{Kashlinsky:2018}. 

We identify the X-ray auto-power from the new sources and discuss prospects for probing it with the forthcoming {\it eROSITA} X-ray mission in conjunction with the source-subtracted CIB measurements from the {\it Euclid}-LIBRAE\footnote{\url{\tiny https://www.euclid.caltech.edu/page/Kashlinsky\%20Team}} project. The reconstructed CXB power from these new sources is so weak that it {\it cannot be isolated directly in the current or forthcoming X-ray missions}. The proposed experimental configuration thus appears uniquely suitable in identifying important information about the BH sources responsible for the observed coherence and their epochs and distribution. After specifying the instrumentational configuration ({\it Euclid}-LIBRAE for CIB and {\it eROSITA} for CXB), we define theoretical expectations and show that, unlike low-$z$ sources, BHs at high $z$ exhibit a well-defined shape of the auto- and cross-power spectrum, which rises at $2\pi/q<0.5^\circ$, robustly peaks at $\sim2^\circ$--$3^\circ$, and traces the Harrison-Zeldovich (HZ) regime, $P\propto q$, at larger angular scales $2\pi/q$. These CXB-CIB cross-powers between {\it source-subtracted} LIBRAE-based CIB and the {\it net} (unclipped) CXB from the X-ray missions are derivable {\it using only harmonics corresponding to their common resolution}. We evaluate the uncertainties in this measurement, which require inputs in the {\it net} X-ray and IR powers for this setup. 
The {\it eROSITA-Euclid} configuration appears currently the most optimal to probe the cross-power 
and the CXB power arising in these BH populations. 
\section{Motivation}
\label{sec:motivation}

\begin{figure}[h!]
\includegraphics[width=6.5in]{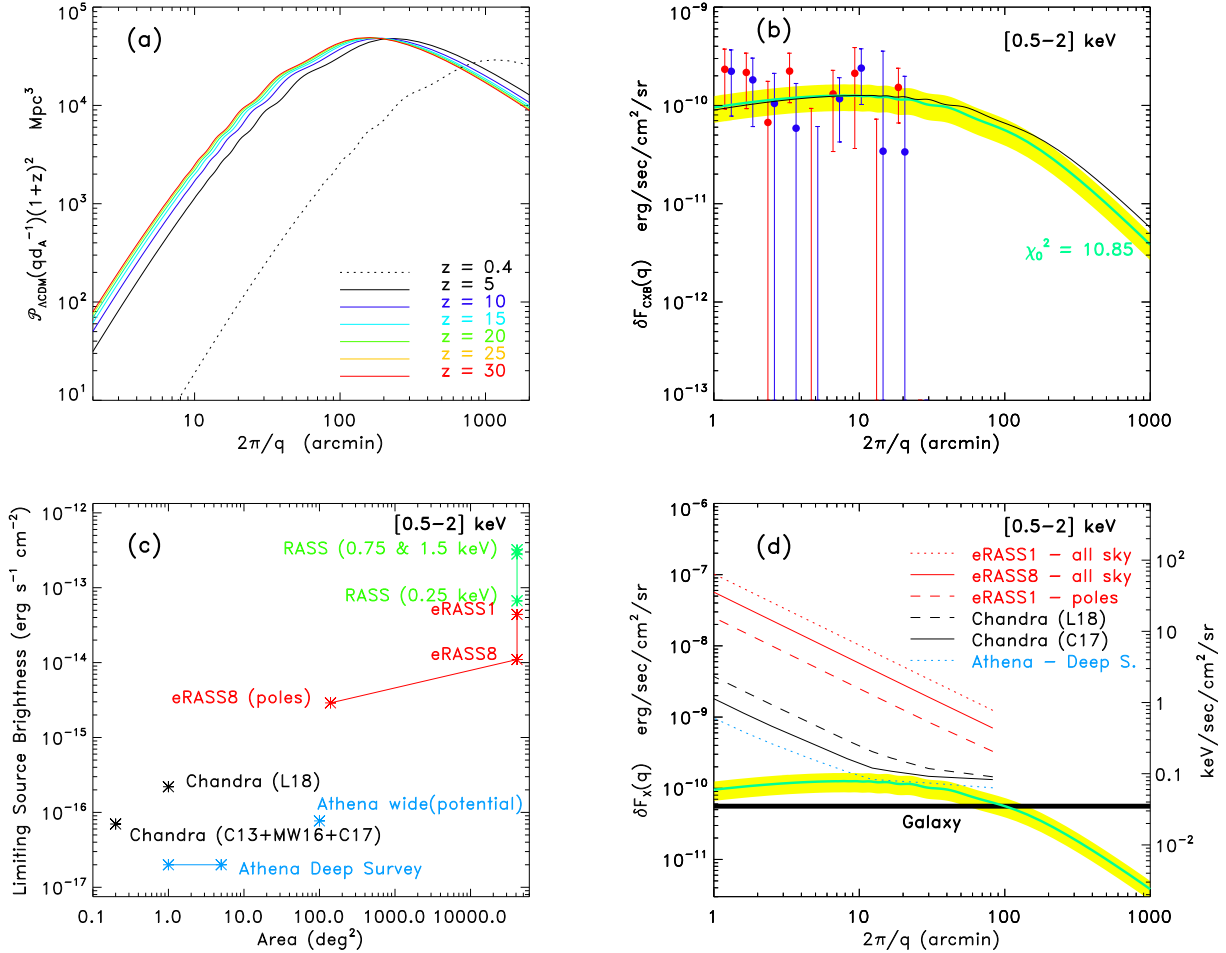}
 \caption[]{(a) Underlying $\Lambda$CDM power spectrum, $ {\cal P}_{\Lambda{\rm CDM}}(q/d_A)$, at $z=0.4,5,10,15,25,30$ projected  to their corresponding distances $d_A=1.1, 5.5, 6.6, 7.2, 7.5, 7.9h^{-1}$ Gpc. At $z\gsim 3$ the Universe is in the Einstein-deSitter regime with density fluctuations growing $\propto (1+z)^{-1}$, so the power is shown multiplied by $(1+z)^2$. (b) Blue (3.6\mic) and red (4.5\mic) circles with errors show $S$ derived using the IRAC/Chandra measurements of \cite{Cappelluti:2017}.  Solid green line is the best fit assuming 1) the power over $1'<2\pi/q<20'$ combining the data at 3.6 and 4.5 \mic\ and 2) the template from (a) projected to $d_A=7$ Gpc ($z=15$); the yellow regions marks 1$\sigma$ deviation of the fit which gives $\chi^2_0=10.85$ for 18 data points (assuming the IRAC bands, with separate optical paths, give independent measurements). Black solid line shows the template projected to $z=5$. (c) X-ray sensitivities. (d) Residual CXB, $\delta F_{\rm X}$, from sources fainter than sensitivity displayed in (c) from with CXB from the new populations identified with the CIB-CXB coherence. Thick horizontal line shows Galactic X-ray fluctuations (extrapolated from
measurements; see Sec.\ref{sec:foregrounds}).}
\label{fig:fig1}
\end{figure}

Source-subtracted CIB fluctuations will be measured in 3 NISP bands from the {\it Euclid} Wide Survey (EWS) covering 15,000 deg$^2$ to AB$\sim 25$ with instantaneous FoV of 0.5 deg$^2$ \citep{Laureijs:2011,Laureijs:2014}. Even excluding areas with substantial Galactic foregrounds, EWS will provide source-subtracted CIB power to sub-percent statistical accuracy. The clipping fraction for CIB maps from Euclid will be $\lsim 10\%$ \citep{Kashlinsky:2018} requiring no significant masking corrections of Fourier amplitudes, $\Delta_1(\vec{q})$. 
Using EWS data, LIBRAE will measure source-subtracted CIB anisotropies on sub-degree angular scales allowing to probe the CIB-CXB cross-power using with the contemporaneous  {\it eROSITA} X-ray mission. {\it eROSITA} will survey the full sky with 15--28$''$ resolution after 6 months (eRASS1), and add repeated coverage over the next 3.5 years to reach $\sigma_X \sim 10^{-14}$erg/s/cm$^2$ (eRASS8). Additional depth is reached at the polar region of 140deg$^2$ \citep{Merloni:2012}. See Fig.\ref{fig:fig1}c.

Cross-power between backgrounds at bands 1 (IR) and 2  (X-ray) is:
\begin{equation}
P_{12} = \int_0^{z_1} \frac{\partial F_1}{\partial z} \frac{\partial F_2}{\partial z} \frac{{\cal P}(qd_A^{-1}(z);z)}{d_A^2 cH^{-1}(z)} dz
\label{eq:limber1}
\end{equation}
where $H(z)=H_0[\Omega_{\rm m}(1+z)^3+\Omega_\Lambda]^{1/2}$, $d_A(z)=c\int_0^z H^{-1}(z)dz$ and $1+z_1=\lambda_1/\lambda_{Ly\alpha}=15(\lambda_1/1.8\mic)$ is the Lyman cutoff for emissions reaching filter at (rest-frame IR) wavelength $\lambda_1$ with $\partial F_1/\partial z$ being the flux rate production at that wavelength (e.g. NISP H-band). For scales subtending linear density field, the 3-D power spectrum is linearly biased with respect to the underlying $\Lambda$CDM power, ${\cal P}(k,z)=b_1(z) b_2(z) {\cal P}_{\Lambda{\rm CDM}}(k,z)$.  In linear regime and at $z\gsim 3$ density perturbations grow so that  $(1+z)^{-2}{\cal P}_{\Lambda{\rm CDM}}(k,z)=$const, which is shown in  Fig. \ref{fig:fig1}a at various $z$. At $z=12$ the comoving distance is $d_A\equiv d_0 \simeq 6.8h^{-1}$Gpc varying by $\sim \pm5\%$ over $9<z<16$, the range defined by the Ly$\alpha$ cutoff at {\it Euclid}/NISP J,H filters. Fig.\ref{fig:fig1}a shows that sources at $z\gsim 5$ exhibit an approximately $z$-independent power spectrum template which  defines robustly the cross- and auto-power shapes from sources at those epochs; i.e. the measured cross-power $P_{12} \propto {\cal P}_{\Lambda{\rm CDM}}(q/d_0)\times$(the weighted measure of CXB and CIB from these sources). The HZ regime, ${\cal P}\propto k$, results in cross- and auto- powers $P\propto q$ and is reached at $2\pi/q\gsim 2^\circ$--$3^\circ$.
At the same time, sources at low(er) $z$ have widely varying shapes of ${\cal P}(q/d_A)$ likely resulting in a broad range of possible shapes for the projected power, which will differ noticeably from the high-$z$ component. Thus measuring  the CIB-CXB cross-power with good accuracy over sufficiently wide angular scales is important in probing/verifying the origin of the signal at high $z$.

The CIB-CXB cross-power can be
used to assess the CXB fluctuations, $[\frac{q^2}{2\pi}P_{{\rm CXB}}]^{1/2}$, from:
\begin{equation}
S\equiv \delta F_{\rm CXB}(q) = \sqrt{\frac{q^2}{2\pi}\frac{P_{12}^2}{P_1}} = \frac{\frac{q^2}{2\pi} P_{12}(q)}{\delta F_{\rm CIB}(q)}.
\label{eq:signal}
\end{equation}
This represents the CXB fluctuations produced directly by the new sources to within the product with the square-root of the underlying CXB-CIB coherence, currently measured to exceed $\sqrt{{\cal C}}\gsim 0.4$; if the CIB is  predominantly BH-produced eq.\ref{eq:signal} would give the true underlying CXB from the new sources with ${\cal C}\simeq1$. 
Because coherence  is $\leq 1$, eq.\ref{eq:signal} sets a lower limit on the CXB power from the sources. 

From the {\it Spitzer/Chandra}-measured cross- and auto-CIB  powers ($P^{\rm IRAC}_{12},P^{\rm IRAC}_{\rm CIB}$) one can derive the CXB power, eq.\ref{eq:signal}, as $S = \frac{q}{\sqrt{2\pi}}P_{12} ^{\rm IRAC}/[P_{\rm CIB} ^{\lambda_{\rm IRAC}}]^{1/2}$. The green line shows the CXB power from the new sources with the normalization of the CIB power using the $\Lambda$CDM template at $d_A=7h^{-1}$Gpc normalized per \cite{Kashlinsky:2015}. 
While the fit is statistically acceptable, the figure shows the limitations of the current measurements for proper interpretation: the signal is probed with limited accuracy and is restricted to $\lsim20^\prime$. 

Fig.\ref{fig:fig1}c shows the depth and angular coverage from X-ray data. Fig.\ref{fig:fig1}d shows $\delta F_{\rm CXB}(q)$  from sources remaining in the data compared to the signal from populations responsible for the measured source-subtracted CIB fluctuations and cross-power with unresolved CXB. We estimated the angular power spectrum from known extragalactic X-ray point sources following \citet{Helgason:2014}. AGNs are sparse and bright; therefore their power spectrum tends to be shot-noise dominated out to $\sim$few degrees \citep{Helgason:2014,Kolodzig:2017}. We adopt the X-ray population model of \citet{Gilli:2007}, in agreement with observed source counts \citep[e.g. ][]{Luo:2017}, and calculate the associated shot-noise power below a given flux limit, neglecting the clustering term from AGN. Galaxies contain high- and low-mass X-ray binaries whose X-ray luminosities scale with star formation rate and stellar mass respectively. For the galaxy population and its clustering, we use a semi-analytic galaxy formation model based on the Millennium simulation \citep{Henriques:2015}, which reproduces the observed star formation history and stellar mass function as a function of redshift. We assign each source its luminosity distance and an X-ray brightness using the $L_X$-SFR$/M_\star$ relation from \citet{Lehmer:2016}, accounting for its scatter. We create a model X-ray image inserting each source in its projected position. Eliminating sources above a given flux limit, the power spectrum is calculated directly from the image using the 2D FFT.

Fig.\ref{fig:fig1}d shows that {\it to robustly probe the new sources in direct CXB measurements, one would need integrations significantly deeper than what will be available while doing this over a large sky area.} 
Hence, the potential of the proposed CIB-CXB cross-power measurement using {\it Euclid} and {\it eROSITA}.

The cross-power between source-subtracted CIB from LIBRAE/Euclid and CXB from {\it eROSITA} can be evaluated as follows: 1) take X-ray diffuse maps with minimal (corresponding to the X-ray survey limits) clipping and keep all harmonics in $\Delta_2(\vec{q})$, 2) take clipped CIB maps and keep only the same harmonics in $\Delta_1(\vec{q})$. Then 3) evaluate the cross-power over $\sigma_0<2\pi/q<20^\circ$ as $P_{12}(q)=\langle |\Delta_1(\vec{q})\Delta^*_2(\vec{q})| H(\frac{2\pi}{q}-\sigma_0)\rangle$, where the CIB power is (presumably) dominated by the new sources. ($H(x)$ is the Heaviside step-function and $\sigma_0\sim30''$ for {\it eROSITA}).   This is similar to the methodology for measuring CMB-CIB cross-power with {\it Euclid} all-sky CIB data to probe the IGM at $z\gsim 10$ (Atrio-Barandela \& Kashlinsky 2014). The known sources will contribute only negligibly to $P_{12}(q)$ as we discuss in Sec.\ref{sec:foregrounds}, but will contribute to the noise on it, eq. \ref{eq:sigma}.  The half-energy-width (HEW) of the {\it eROSITA} PSF is 28$\arcsec$ with extended tails, so mask leakage from bright sources is a  potential noise source.  
This effect was estimated as subdominant in  \citep[][Sec. VII.D]{Kashlinsky:2018} at small angular 
scales, becoming less important at the larger scales. Conservatively we take the smallest angular scale to be $\sim 1'$ for the CXB-CIB cross-power estimates, where also CIB contributions from remaining known sources are smaller than the expected high-$z$ component.

In the presence of X-ray and IR maps with net diffuse light power $P_{\rm CIB}(q),P_{\rm CXB}(q)$ the error on the measured cross-power at $q$ from an ensemble of $n_q$ independent Fourier elements is
\begin{equation}
\sigma _{\rm IR,X}=\sqrt{\frac{P_{\rm IR}P_{\rm X}}{2n_q}}
\label{eq:sigma}
\end{equation}
The cosmological cross-power is expressable in terms of the cosmological auto-powers as $P_{12}(q) = \sqrt{{\cal C}(q) P_{\rm CIB}(q)P_{\rm CXB}(q)}$ with the coherence potentially reaching ${\cal C}\sim1$. The signal-to-noise of the cross-power measurements, for one single patch of size $\Theta_{\rm patch}$ on the side, is
\begin{equation}
\frac{S}{N}|_{\rm patch}=[2n_q]^{1/2}\left[\frac{\delta F_{\rm CXB}(q)}{\delta F_{\rm X}(q)}\right] \left[\frac{\delta F_{\rm CIB}(q)}{\delta F_{\rm IR}(q)}\right]
\label{eq:s2n}
\end{equation}
where $\delta F_{\rm X}=[q^2P_{\rm X}/(2\pi)]^{1/2}$ is the net diffuse flux fluctuation in the X-ray maps from sources remaining at the X-ray depth and Galaxy, $\delta F_{\rm CIB}$ is defined similarly at its own IR depth. The number of elements, $n_q$, for the patch which goes into determining the power at each $q$ depends on the  patch area, $\Theta_{\rm patch}^2$, and the $q$-binning. We write $n_q\simeq \pi (\frac{\Theta_{\rm patch}}{2\pi/q})^2 (\frac{\Delta q}{q})$, where $\Delta q$ is the bin-width over which the power at the given wavenumber $q$ is averaged in the Fourier plane.
The direct measurement of $\delta F_{\rm CXB}$ [Eq. (\ref{eq:signal})]  would have a corresponding $
\frac{S}{N}|_{\rm direct}=[n_q]^{1/2}\left[\frac{\delta F_{\rm CXB}(q)}{\delta F_{\rm X}(q)}\right]^2
$. 
Employing the cross power instead of direct measurement of the CXB fluctuations leads to improved $S/N$ when $
(2{\cal C})^{1/2} [\delta F_{\rm CIB}(q)/\delta F_{\rm IR}(q)]>\delta F_{\rm CXB}(q)/\delta F_{\rm X}(q)
$.
Thus improvements are made when the total background is more strongly influenced by the cosmological component in the IR emission
than in the X-ray emission, and as long as a low coherence does not counteract the benefit of using cleaner IR data: the method identifies the new CXB contributing populations if 1) they are strongly coherent with the CIB sources, 2) the CIB power of the new sources is isolated, but 3) their CXB contributions are drowned in the noise and other X-ray sources. Furthermore, a given $S/N$ Gaussian-distributed cross-power corresponds to higher confidence levels than the $\chi^2$-distributed auto-power. 

The net signal-to-noise over a wide net area $A$ covering $A/\Theta_{\rm patch}^2$ such patches becomes:
\begin{equation}
\left(\frac{S}{N}\right)^2=\sum_{\rm patches} 2n_q\left[\frac{\delta F_{\rm CXB}(q)}{\delta F_{\rm X}(q)}\right] ^2\left[\frac{\delta F_{\rm CIB}(q)}{\delta F_{\rm IR}(q)}\right] ^2\equiv 2n_q W(q)
\label{eq:s2n_final}
\end{equation}
where $W(q)\equiv \sum_{\rm patches} \left[\frac{\delta F_{\rm CXB}(q)}{\delta F_{\rm X}(q)}\right] ^2\left[\frac{\delta F_{\rm CIB}(q)}{\delta F_{\rm IR}(q)}\right] ^2$ is evaluated in Sec. \ref{sec:foregrounds}; this last expression being appropriate when analysis of a large area of sky is performed after dividing it into smaller patches.  $W\propto A$ when the terms inside the sum for $W$ are $q$-independent. 

\section{CXB-CIB cross-power uncertainties}
\label{sec:foregrounds}

\begin{figure}[h!]
   \centering
   \includegraphics[width=6.5in]{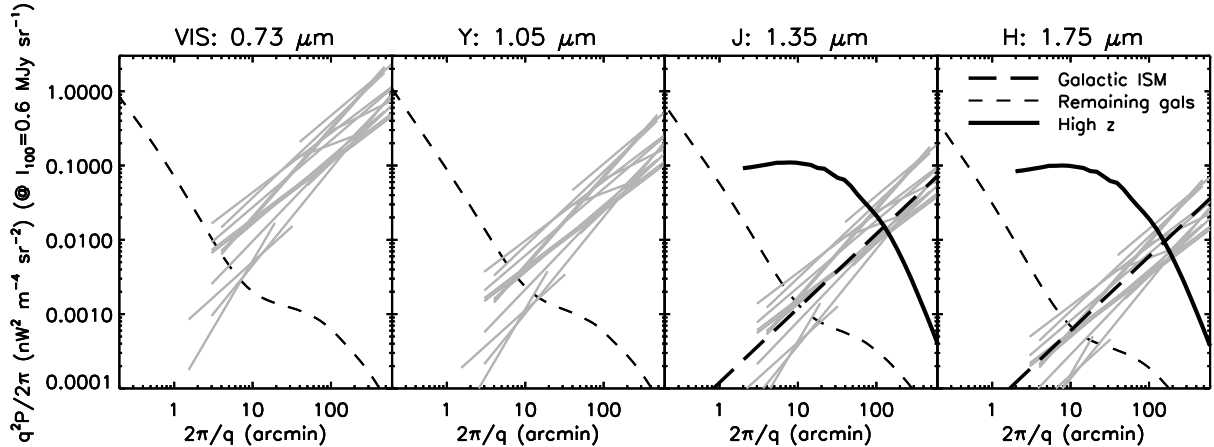} 
   \caption{Normalized cirrus power spectra (grey power laws) 
   from various published studies
   translated to the {\it Euclid} wavelengths. Legend marks IR components adopted in subsequent $S/N$ estimates: 1) the contribution from remaining known galaxies, 2) the Galactic ISM/cirrus, and 3) the $\Lambda$CDM power spectrum of the modeled high-$z$ CIB normalized to {\it Spitzer} CIB data. 
}   \label{fig:power_galaxy}
\end{figure}

To evaluate the $S/N$ one needs the 
ratios of the cosmic background to total powers in each of the IR and X-ray bands. For CIB, we adopt a theoretical model based on the IMF500 model from \cite{Helgason:2016}, which fits the {\it Spitzer} excess CIB measurements as discussed in \cite{Kashlinsky:2015} with the mean formation efficiency per halo of $f_\star=0.04$ ending at $z_{\rm end}=10$. This model, shown in Fig. \ref{fig:power_galaxy}, has negligible CIB contributions in the NISP Y filter, but dominates remaining known galaxy contributions at J and H. Strictly speaking it corresponds to stellar emissions from very massive stars ($500M_\odot$ each), but, because those radiate at the Eddington limit with $L\propto M$ as do BHs, can be straightforwardly rescaled to BH emissions. 

In deep IR images, source subtraction removes Galactic stars and 
resolved extragalactic sources. At small angular scales (e.g.\ $<20''$) or
for relatively shallow observations, the shot noise of the remaining faint 
galaxies of known populations will still contribute to the overall IR background
power. Emission from the Galactic ISM will also remain, and 
provide a potentially dominant foreground to the CIB. The power spectrum of Galactic ISM emission has 
been measured at 100 $\mu$m \citep[e.g. ][]{Gautier:1992,Wright:1998,Miville-Deschenes:2002,Kiss:2003,Lagache:2007,Bracco:2011,Penin:2012}. 
These studies were in regions of various ISM brightness, but we empirically find using the data from these studies 
that $P(100') \propto \langle I_{100} \rangle^{2.7}$, and renormalize all the
reported power spectra to a mean ISM 100 $\mu$m intensity of 0.6 MJy/sr.
We 
rescaled the power spectra from 100$\mu$m to near-IR 
using a typical 2--300$\mu$m spectrum of the 
ISM\footnote{\url{https://irsa.ipac.caltech.edu/data/SPITZER/docs/files/spitzer/background.pdf}}, 
extended to shorter wavelengths using the diffuse
Galactic light measurements of \cite{Brandt:2012},  and adjusting at  
$<4$\mic\ to match the redder 3.6/100 $\mu$m color \citep{Arendt:2003}. 
The resultant ISM power spectra at different Euclid wavelengths are shown in 
Fig.\ref{fig:power_galaxy}, along with the nominal $P \propto (2\pi/q)^3$ power law that we use
to represent the ISM in further calculations, i.e. $P \propto q^{-3}\langle I_{100} \rangle^{2.7}$.

Contributions from remaining known galaxies, dashes in Fig.\ref{fig:power_galaxy} \citep[see Sec. VII.C, ][]{Kashlinsky:2018}, are small compared to the model CIB at the scales and wavelengths of interest, $2\pi/q>1'$ and J,H. Because of the much deeper IR threshold of removing individual sources in EWS than in {\it eROSITA}, they contribute only negligibly to the IR-X cross-power. We evaluated the cross-power from known sources remaining at the EWS H-band projected depth using the methodology described in Sec.\ref{sec:motivation}. The known galaxies were found to contribute $q^2P_{12}/(2\pi)\simeq 10^{-12}$(erg/sec/cm$^2$/sr$\cdot$nW/m$^2$/sr) at $2\pi/q\sim5'$, dropping to $\sim2\times 10^{-13}$ at $\sim 30'$ while reaching $10^{-11}$ at $2\pi/q=1'$. The CIB-CXB fluctuations are measured at 3.6\mic\ in \citet[][Fig. 2 there]{Cappelluti:2017}  to be $\simeq 10^{-11}$ in these units leading to numbers displayed in Fig.\ref{fig:fig1}b,d. The H-band CIB-CXB cross-power from the new sources would be {\it larger} by the corresponding ratio of the CIB powers at the H to 3.6\mic\ bands or a factor of $\sim4$ per Fig.\ref{fig:power_galaxy},right, assuming the coherence remains constant. This component is thus neglected in the overall budget of $S/N$.


The ratio of the amplitudes of the CXB and the total X-ray fluctuations is estimated from the components shown in Fig.\ref{fig:fig1}d. The total 
X-ray power is the sum of the CXB, the remaining extragalactic component for
any given survey, and a Galactic ISM component.
The high latitude X-ray power has been measured by \cite{Sliwa:2001} using
{\it ROSAT}, and over a more limited region using much deeper
{\it Chandra} data by \cite{Kolodzig:2018}. The largest scale measurements 
(at $\sim 2\arcdeg$) by \cite{Kolodzig:2018} can be extrapolated (using 
$P_{\rm X} \propto q^2$ or $\delta F_{\rm X}$=const) as an upper limit on 
ISM X-ray emission (see Fig.\ref{fig:fig1}d), consistent with the largest 
scale {\it ROSAT}/RASS measurements. 
Comparison with the expected CXB power spectrum reveals that at high latitudes, 
the ISM power will be comparable to or fainter than the extragalactic power. 

\begin{figure}[h!]
   \centering
   \includegraphics[width=3.15in]{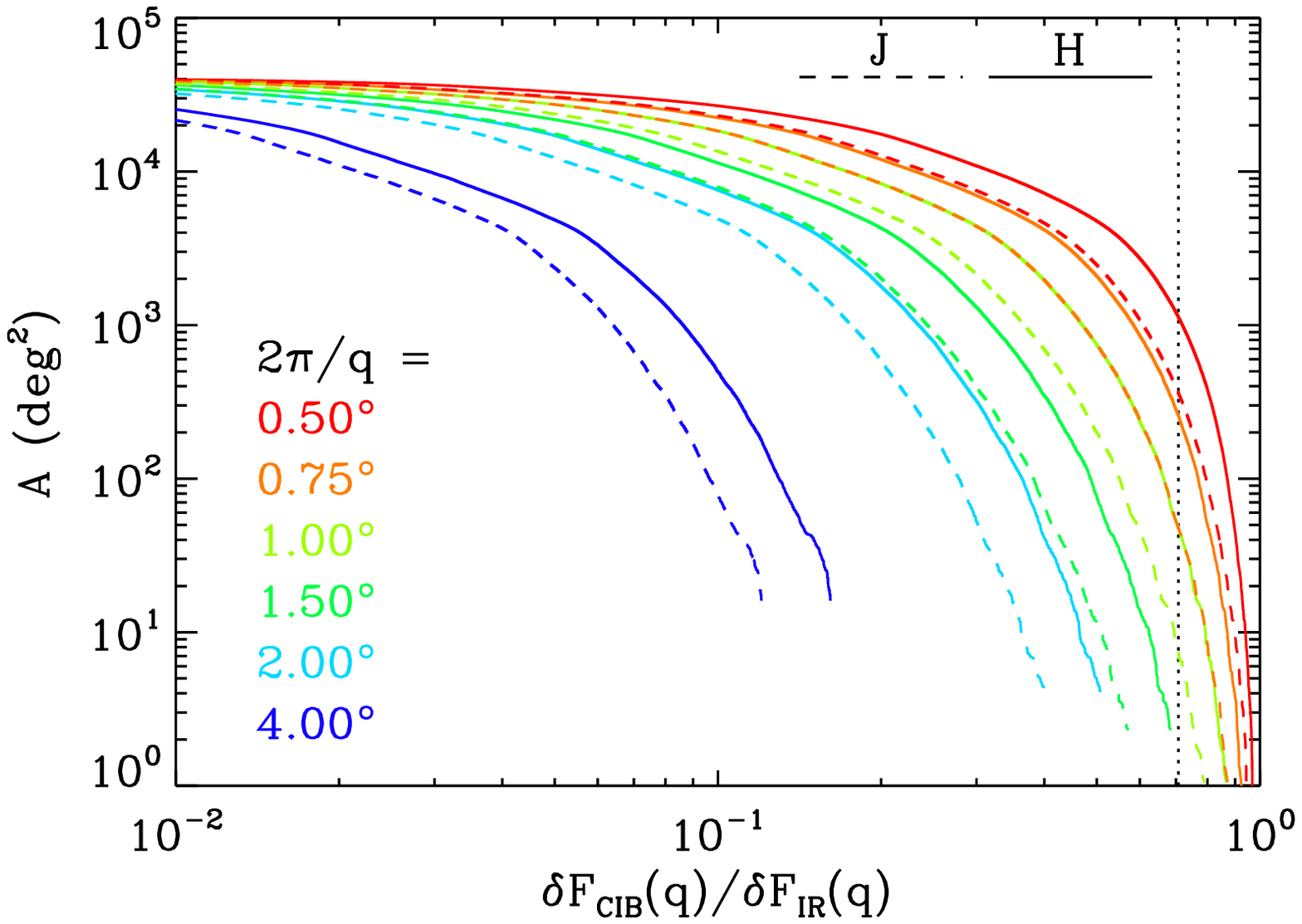} 
   \includegraphics[width=3.15in]{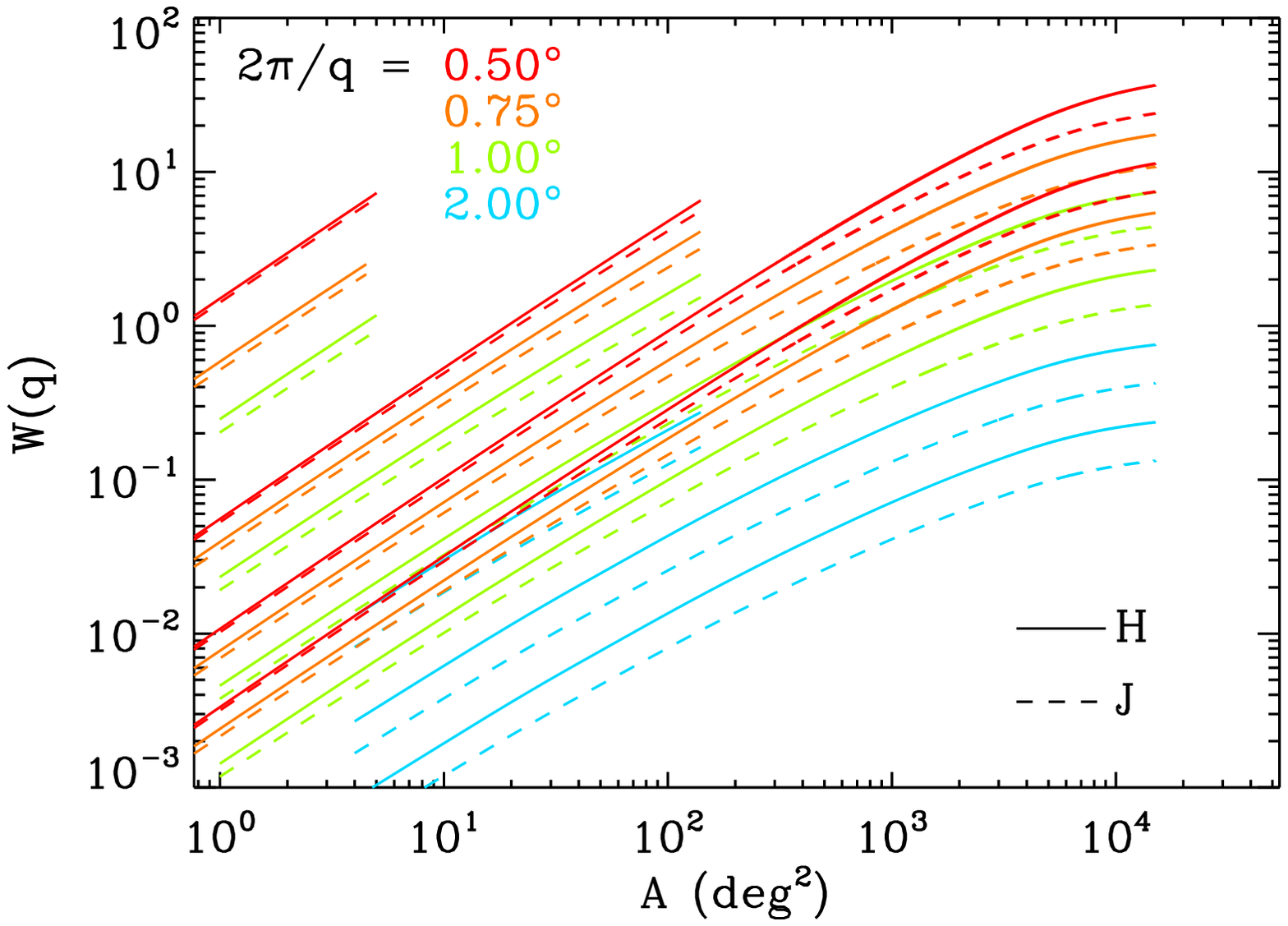} 
   \caption{Left: The area over which 
   $\delta F_{\rm CIB}(q)/\delta F_{\rm IR}(q)$ exceeds a given value
   as a function of $2\pi/q$ and IR band. The expected contributions of 
   emission from the Galactic ISM limits that area of the sky where the 
   CIB can be well measured, especially at large $2\pi/q$.
Dotted vertical line shows $\delta F_{\rm CIB}(q)/\delta F_{\rm IR}(q) = \sqrt{0.5}$
   identifying where $P_{\rm CIB} > 0.5 P_{\rm IR}$. 
   Right: $\delta F_{\rm CIB}(q)/\delta F_{\rm IR}(q)$ is combined with 
   comparable X-ray evaluations to estimate $W(q) \propto (S/N)^2$
   expected for IR--X-ray cross power measurements 
made over increasingly larger areas of the 
   sky, $A$. 
   $W(q) \propto A$ up to $A\sim 10^2$ deg$^2$, but as larger areas are affected by 
   higher foregrounds, the $S/N$ accumulates more slowly. 
   The curves limited to $A < 5$ and 140 deg$^2$ represent the Athena-Deep
   and eRASS poles surveys. The eRASS1 and eRASS8 surveys cover the entire 
   EWS area, with the deeper 
   eRASS8 providing higher $S/N$.} 
   \label{fig:s2n}
\end{figure}

%
The final consideration in evaluating the $S/N$ of the prospective measurement involves $n_q$ in the region of measurements. The ratios
${P_{\rm CIB}}/{P_{\rm IR}}$ and ${P_{\rm CXB}}/{P_{\rm X}}$ discussed above
are representative for high latitude regions. For regions closer to the Galactic
plane, the foreground emission increases relative to the cosmic background. 
The ratios and thus the $S/N$ for such regions decrease. To estimate this 
effect, we made maps of the amplitudes of $P_{\rm IR}$ and $P_{\rm X}$,
by rescaling the DIRBE 100 $\mu$m and {\it ROSAT} R4+R7 maps. Thus the $S/N$ can be evaluated by summing over increasingly 
larger areas (increasing $n_q$ and/or the number of patches), 
but decreasingly smaller ratios of 
background to total power. The resultant $W$'s are shown in 
Fig.\ref{fig:s2n},right. As one approaches large areas, the value of $\sqrt{W}$, governing the overall achievable $S/N$, saturates and increasing the net area leads to progressively lower benefits as $A\gsim$ a few thousand deg$^2$.
\section{Prospects from upcoming X-ray missions}
\label{sec:summary}

\begin{figure}[h!]
\includegraphics[width=4.5in]{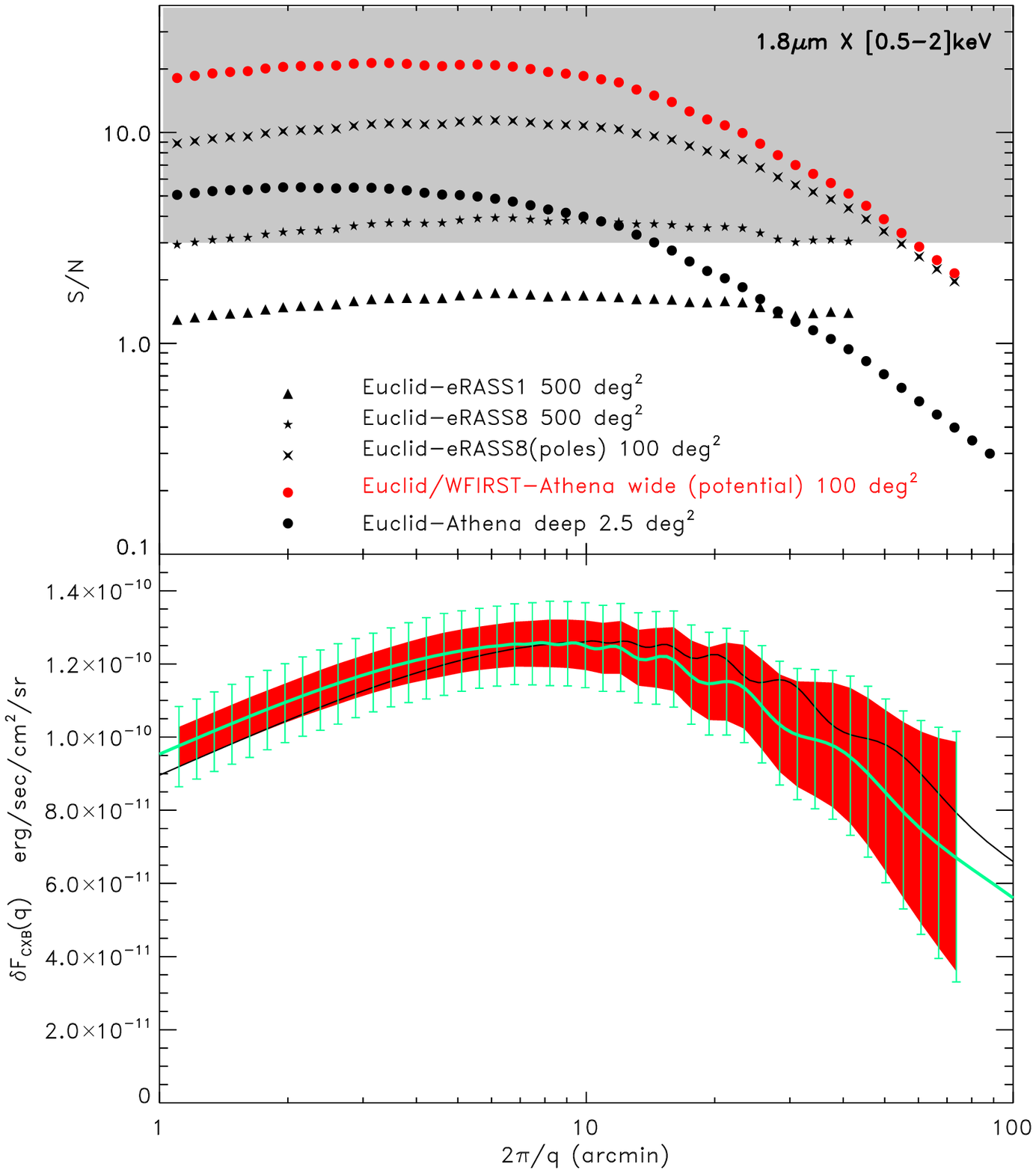}
 \caption[]{Top: Overall S/N with $\Delta q/q=0.1$ for NISP's H-band CIB for the marked configurations; for J-band the numbers would be similar and are not shown for brevity. Shaded area marks $S/N \geq 3$. 
Bottom: the expected signal in Fig. \ref{fig:fig1}d if generated by sources at $z=15$ (green solid) and $z=5$ (black dotted); in reality the cross-power is a weighted sum/integral over $z$. Vertical bars correspond to 1-$\sigma$ uncertainties projected in the eROSITA Poles survey covering $A$=100 deg$^2$ and $\Delta q/q=0.1$.
Red shows the 1-$\sigma$ limits for the same configuration with a potential Athena survey. 
}
\label{fig:fig3}
\end{figure}
We now evaluate the $S/N$ of the aforementioned configurations using the EWS CIB. The dotted line in Fig.\ref{fig:s2n},left demarcates where CIB and foregrounds contributions to the net IR power become comparable: $P_{\rm CIB}$ becomes dominant over the IGM/cirrus contributions over 
1) $A\sim 500$ deg$^2$ area if square patches of 0.7$^\circ$ on the side are selected (comparable to NISP's FoV), 2) $A\sim 100$ deg$^2$ if the area $A$ is probed in square patches of 1 deg$^2$, or $2\pi/q\lsim 1^\circ$, 
and 3) $A\sim 2.5$ deg$^2$ if probing out to $2\pi/q\sim 1.5^\circ$. These configurations correspond  to 1) eROSITA1/8, 2) eROSITA-poles, and 3) Athena-deep and we estimate the resultant $S/N$ there. This is a conservative limit as covering more area would result in a better signal out to $S/N\simeq \sqrt{2n_q\times {\rm max}[W(q)]}$; Fig.\ref{fig:s2n},right shows that $\gsim50\%$ of $S/N$ for eRASS all-sky configurations is reached at $A\lsim 1,000$ deg$^2$. In estimating $S/N$ we adopted Fourier binning of $\Delta q/q=0.1$ and assumed 
that the patches analyzed are of the same extent as the scale
of interest. If the patches analyzed are much larger than the scales measured, 
these results can be binned over wider $\Delta q$, which
increases $n_q\propto \Delta q$ and $S/N\propto \sqrt{\Delta q}$. 

Fig.\ref{fig:fig3},top shows the achievable $S/N$ for the NISP H-band. As Fig. \ref{fig:power_galaxy} indicates the numbers using J-band CIB would be similar, although combining the two IR bands would not lead to appreciable increase of the overall S/N because the uncertainties are due to populations remaining in the data which correlate between the IR bands, rather than instrumental noise. {\it eROSITA} appears to be very well suited for this measurement at least as far as the soft X-ray band is concerned, with the best configuration being the eROSITA-poles survey.
At a given X-ray depth $S/N \propto A^{1/2}(\Delta q/q)^{1/2}$. $S/N\geq 3$ at each scale can be achieved already with the eROSITA1 maps with $\Delta q/q \simeq 0.25 (A/1,000 {\rm deg}^2)^{-1}$ out to $\sim 1^\circ$, enabling a good first look at the cross-power. Athena-Deep would have good sensitivity out to $\sim 30^\prime$, but because of its small area will lose sensitivity at larger scales. eRASS8 and eROSITA-Poles surveys will allow probing the cross-power highly accurately out to $\sim 2^\circ$. 
We note that, given the high $S/N$, masking corrections may be important to make when doing FFTs, unless avoided by the CPU-time intensive correlation function analysis.

The shape of the cross-power, as probed by the eROSITA-poles, could potentially probe the epochs of the new sources although detailed studies of the masking effects at such high $S/N$ would be required to answer this quantitatively as we are now conducting. However, the high-$z$ origin of the signal would be probed directly with the Lyman-break cutoff from the EWS VIS and NISP cross-powers \citep{Kashlinsky:2018}.

Post-{\it Euclid/eROSITA}, {\it WFIRST} to-be-launched in late 2020's will cover $\sim$2,000 deg$^2$ at 4 NIR bands to deeper levels than EWS \citep{Spergel:2015} and ESA's {\it Athena} X-ray mission launch \citep{Nandra:2013} is planned in early 2030s. Athena  with  $\sim5''$ HEW PSF and a collecting area much larger than {\it eROSITA}'s, will be a much more sensitive telescope for faint diffuse emission to study the fluctuations over a broader range of angular scales.
An {\it Athena} wide survey  (say, $A=100$deg$^2$) at depths comparable to the Chandra integrations used in \cite{Cappelluti:2017} will achieve the $S/N$ shown with red in Fig.\ref{fig:fig3}. This exceeds {\it eROSITA} and would require only as much integration as the planned {\it Athena} Deep Survey ($A\simeq$2--5 deg$^2$), assuming $A\sigma_X^2=$const.
 Such an {\it Athena} wide survey would probe the CIB-CXB cross-power, and the corresponding CIB auto-power, at the levels where remaining CXB comes very close to that from the new sources (Fig.\ref{fig:fig1}d). In this situation  one would be able to probe the {\it intrinsic} coherence due to BHs among the CIB sources.

We acknowledge support from NASA/12-EUCLID11-0003 ``LIBRAE: Looking at Infrared Background Radiation Anisotropies with Euclid" project (\url{http://librae.ssaihq.com}), NASA award \#80GSFC17M0002 (RGA), and Icelandic Research Fund grant \#173728-051 (KH). 
%


\begin{thebibliography}{}
\expandafter\ifx\csname natexlab\endcsname\relax\def\natexlab#1{#1}\fi

\bibitem[{{Arendt} \& {Dwek}(2003)}]{Arendt:2003}
{Arendt}, R.~G., \& {Dwek}, E. 2003, \apj, 585, 305

\bibitem[{{Arendt} {et~al.}(2010){Arendt}, {Kashlinsky}, {Moseley}, \&
  {Mather}}]{Arendt:2010}
{Arendt}, R.~G., {Kashlinsky}, A., {Moseley}, S.~H., \& {Mather}, J. 2010,
  \apjs, 186, 10

\bibitem[{{Bracco} {et~al.}(2011){Bracco}, {Cooray}, {Veneziani}, {Amblard},
  {Serra}, {Wardlow}, {Thompson}, {White}, {Auld}, {Baes}, {Bertoldi},
  {Buttiglione}, {Cava}, {Clements}, {Dariush}, {de Zotti}, {Dunne}, {Dye},
  {Eales}, {Fritz}, {Gomez}, {Hopwood}, {Ibar}, {Ivison}, {Jarvis}, {Lagache},
  {Lee}, {Leeuw}, {Maddox}, {Micha{\l}owski}, {Pearson}, {Pohlen}, {Rigby},
  {Rodighiero}, {Smith}, {Temi}, {Vaccari}, \& {van der Werf}}]{Bracco:2011}
{Bracco}, A., {Cooray}, A., {Veneziani}, M., {et~al.} 2011, \mnras, 412, 1151

\bibitem[{{Brandt} \& {Draine}(2012)}]{Brandt:2012}
{Brandt}, T.~D., \& {Draine}, B.~T. 2012, \apj, 744, 129

\bibitem[{{Cappelluti} {et~al.}(2013){Cappelluti}, {Kashlinsky}, {Arendt},
  {Comastri}, {Fazio}, {Finoguenov}, {Hasinger}, {Mather}, {Miyaji}, \&
  {Moseley}}]{Cappelluti:2013}
{Cappelluti}, N., {Kashlinsky}, A., {Arendt}, R.~G., {et~al.} 2013, \apj, 769,
  68

\bibitem[{{Cappelluti} {et~al.}(2017){Cappelluti}, {Arendt}, {Kashlinsky},
  {Li}, {Hasinger}, {Helgason}, {Urry}, {Natarajan}, \&
  {Finoguenov}}]{Cappelluti:2017}
{Cappelluti}, N., {Arendt}, R., {Kashlinsky}, A., {et~al.} 2017, \apjl, 847,
  L11

\bibitem[{{Cooray} {et~al.}(2004){Cooray}, {Bock}, {Keatin}, {Lange}, \&
  {Matsumoto}}]{Cooray:2004}
{Cooray}, A., {Bock}, J.~J., {Keatin}, B., {Lange}, A.~E., \& {Matsumoto}, T.
  2004, \apj, 606, 611

\bibitem[{{Cooray} {et~al.}(2012){Cooray}, {Smidt}, {de Bernardis}, {Gong},
  {Stern}, {Ashby}, {Eisenhardt}, {Frazer}, {Gonzalez}, {Kochanek},
  {Koz{\l}owski}, \& {Wright}}]{Cooray:2012}
{Cooray}, A., {Smidt}, J., {de Bernardis}, F., {et~al.} 2012, \nat, 490, 514

\bibitem[{{Gautier} {et~al.}(1992){Gautier}, {Boulanger}, {Perault}, \&
  {Puget}}]{Gautier:1992}
{Gautier}, T.~N., I., {Boulanger}, F., {Perault}, M., \& {Puget}, J.~L. 1992,
  \aj, 103, 1313

\bibitem[{{Gilli} {et~al.}(2007){Gilli}, {Comastri}, \&
  {Hasinger}}]{Gilli:2007}
{Gilli}, R., {Comastri}, A., \& {Hasinger}, G. 2007, \aap, 463, 79

\bibitem[{{Helgason} {et~al.}(2014){Helgason}, {Cappelluti}, {Hasinger},
  {Kashlinsky}, \& {Ricotti}}]{Helgason:2014}
{Helgason}, K., {Cappelluti}, N., {Hasinger}, G., {Kashlinsky}, A., \&
  {Ricotti}, M. 2014, \apj, 785, 38

\bibitem[{{Helgason} {et~al.}(2012){Helgason}, {Ricotti}, \&
  {Kashlinsky}}]{Helgason:2012a}
{Helgason}, K., {Ricotti}, M., \& {Kashlinsky}, A. 2012, \apj, 752, 113

\bibitem[{{Helgason} {et~al.}(2016){Helgason}, {Ricotti}, {Kashlinsky}, \&
  {Bromm}}]{Helgason:2016}
{Helgason}, K., {Ricotti}, M., {Kashlinsky}, A., \& {Bromm}, V. 2016, \mnras,
  455, 282

\bibitem[{{Henriques} {et~al.}(2015){Henriques}, {White}, {Thomas}, {Angulo},
  {Guo}, {Lemson}, {Springel}, \& {Overzier}}]{Henriques:2015}
{Henriques}, B.~M.~B., {White}, S.~D.~M., {Thomas}, P.~A., {et~al.} 2015,
  \mnras, 451, 2663

\bibitem[{{Kashlinsky}(2005)}]{Kashlinsky:2005}
{Kashlinsky}, A. 2005, \physrep, 409, 361

\bibitem[{{Kashlinsky}(2016)}]{Kashlinsky:2016}
---. 2016, \apjl, 823, L25

\bibitem[{{Kashlinsky} {et~al.}(2004){Kashlinsky}, {Arendt}, {Gardner},
  {Mather}, \& {Moseley}}]{Kashlinsky:2004}
{Kashlinsky}, A., {Arendt}, R., {Gardner}, J.~P., {Mather}, J.~C., \&
  {Moseley}, S.~H. 2004, \apj, 608, 1

\bibitem[{{Kashlinsky} {et~al.}(2012){Kashlinsky}, {Arendt}, {Ashby}, {Fazio},
  {Mather}, \& {Moseley}}]{Kashlinsky:2012}
{Kashlinsky}, A., {Arendt}, R.~G., {Ashby}, M.~L.~N., {et~al.} 2012, \apj, 753,
  63

\bibitem[{{Kashlinsky} {et~al.}(2018){Kashlinsky}, {Arendt}, {Atrio-Barandela},
  {Cappelluti}, {Ferrara}, \& {Hasinger}}]{Kashlinsky:2018}
{Kashlinsky}, A., {Arendt}, R.~G., {Atrio-Barandela}, F., {et~al.} 2018,
  Reviews of Modern Physics, 90, 025006

\bibitem[{{Kashlinsky} {et~al.}(2015){Kashlinsky}, {Arendt}, {Atrio-Barandela},
  \& {Helgason}}]{Kashlinsky:2015}
{Kashlinsky}, A., {Arendt}, R.~G., {Atrio-Barandela}, F., \& {Helgason}, K.
  2015, \apjl, 813, L12

\bibitem[{{Kashlinsky} {et~al.}(2005){Kashlinsky}, {Arendt}, {Mather}, \&
  {Moseley}}]{Kashlinsky:2005a}
{Kashlinsky}, A., {Arendt}, R.~G., {Mather}, J., \& {Moseley}, S.~H. 2005,
  \nat, 438, 45

\bibitem[{{Kashlinsky} {et~al.}(2007){Kashlinsky}, {Arendt}, {Mather}, \&
  {Moseley}}]{Kashlinsky:2007a}
---. 2007, \apjl, 654, L5

\bibitem[{{Kiss} {et~al.}(2003){Kiss}, {{\'A}brah{\'a}m}, {Klaas}, {Lemke},
  {H{\'e}raudeau}, {del Burgo}, \& {Herbstmeier}}]{Kiss:2003}
{Kiss}, C., {{\'A}brah{\'a}m}, P., {Klaas}, U., {et~al.} 2003, \aap, 399, 177

\bibitem[{{Kolodzig} {et~al.}(2017){Kolodzig}, {Gilfanov}, {H{\"u}tsi}, \&
  {Sunyaev}}]{Kolodzig:2017}
{Kolodzig}, A., {Gilfanov}, M., {H{\"u}tsi}, G., \& {Sunyaev}, R. 2017, \mnras,
  466, 3035

\bibitem[{{Kolodzig} {et~al.}(2018){Kolodzig}, {Gilfanov}, {H{\"u}tsi}, \&
  {Sunyaev}}]{Kolodzig:2018}
---. 2018, \mnras, 473, 4653

\bibitem[{{Lagache} {et~al.}(2007){Lagache}, {Bavouzet}, {Fernandez-Conde},
  {Ponthieu}, {Rodet}, {Dole}, {Miville- Desch{\^e}nes}, \&
  {Puget}}]{Lagache:2007}
{Lagache}, G., {Bavouzet}, N., {Fernandez-Conde}, N., {et~al.} 2007, \apj, 665,
  L89

\bibitem[{{Laureijs} {et~al.}(2011){Laureijs}, {Amiaux}, {Arduini},
  {Augu{\`e}res}, {Brinchmann}, {Cole}, {Cropper}, {Dabin}, {Duvet}, {Ealet},
  \& et~al.}]{Laureijs:2011}
{Laureijs}, R., {Amiaux}, J., {Arduini}, S., {et~al.} 2011, ArXiv e-prints,
  arXiv:1110.3193

\bibitem[{{Laureijs} {et~al.}(2014){Laureijs}, {Racca}, {Stagnaro},
  {Salvignol}, {Lorenzo Alvarez}, {Saavedra Criado}, {Gaspar Venancio},
  {Short}, {Strada}, {Colombo}, {Buenadicha}, {Hoar}, {Kohley}, {Vavrek},
  {Mellier}, {Berthe}, {Amiaux}, {Cropper}, {Niemi}, {Pottinger}, {Ealet},
  {Jahnke}, {Maciaszek}, {Pasian}, {Sauvage}, {Wachter}, {Israelsson},
  {Holmes}, {Seiffert}, {Cazaubiel}, {Anselmi}, \& {Musi}}]{Laureijs:2014}
{Laureijs}, R., {Racca}, G., {Stagnaro}, L., {et~al.} 2014, in \procspie, Vol.
  9143, Space Telescopes and Instrumentation 2014: Optical, Infrared, and
  Millimeter Wave, 91430H

\bibitem[{{Lehmer} {et~al.}(2016){Lehmer}, {Basu-Zych}, {Mineo}, {Brandt},
  {Eufrasio}, {Fragos}, {Hornschemeier}, {Luo}, {Xue}, {Bauer}, {Gilfanov},
  {Ranalli}, {Schneider}, {Shemmer}, {Tozzi}, {Trump}, {Vignali}, {Wang},
  {Yukita}, \& {Zezas}}]{Lehmer:2016}
{Lehmer}, B.~D., {Basu-Zych}, A.~R., {Mineo}, S., {et~al.} 2016, \apj, 825, 7

\bibitem[{{Li} {et~al.}(2018){Li}, {Cappelluti}, {Arendt}, {Hasinger},
  {Kashlinsky}, \& {Helgason}}]{Li:2018}
{Li}, Y., {Cappelluti}, N., {Arendt}, R.~G., {et~al.} 2018, \apj, 864, 141

\bibitem[{{Luo} {et~al.}(2017){Luo}, {Brandt}, {Xue}, {Lehmer}, {Alexander},
  {Bauer}, {Vito}, {Yang}, {Basu-Zych}, {Comastri}, {Gilli}, {Gu},
  {Hornschemeier}, {Koekemoer}, {Liu}, {Mainieri}, {Paolillo}, {Ranalli},
  {Rosati}, {Schneider}, {Shemmer}, {Smail}, {Sun}, {Tozzi}, {Vignali}, \&
  {Wang}}]{Luo:2017}
{Luo}, B., {Brandt}, W.~N., {Xue}, Y.~Q., {et~al.} 2017, \apjs, 228, 2

\bibitem[{{Merloni} {et~al.}(2012){Merloni}, {Predehl}, {Becker},
  {B{\"o}hringer}, {Boller}, {Brunner}, {Brusa}, {Dennerl}, {Freyberg},
  {Friedrich}, {Georgakakis}, {Haberl}, {Hasinger}, {Meidinger}, {Mohr},
  {Nandra}, {Rau}, {Reiprich}, {Robrade}, {Salvato}, {Santangelo}, {Sasaki},
  {Schwope}, {Wilms}, \& {German eROSITA Consortium}}]{Merloni:2012}
{Merloni}, A., {Predehl}, P., {Becker}, W., {et~al.} 2012, ArXiv e-prints,
  arXiv:1209.3114

\bibitem[{{Mitchell-Wynne} {et~al.}(2016){Mitchell-Wynne}, {Cooray}, {Xue},
  {Luo}, {Brandt}, \& {Koekemoer}}]{Mitchell-Wynne:2016}
{Mitchell-Wynne}, K., {Cooray}, A., {Xue}, Y., {et~al.} 2016, \apj, 832, 104

\bibitem[{{Miville-Desch{\^e}nes} {et~al.}(2002){Miville-Desch{\^e}nes},
  {Lagache}, \& {Puget}}]{Miville-Deschenes:2002}
{Miville-Desch{\^e}nes}, M.~A., {Lagache}, G., \& {Puget}, J.~L. 2002, \aap,
  393, 749

\bibitem[{{Nandra} {et~al.}(2013){Nandra}, {Barret}, {Barcons}, {Fabian}, {den
  Herder}, {Piro}, {Watson}, {Adami}, {Aird}, {Afonso}, \&
  et~al.}]{Nandra:2013}
{Nandra}, K., {Barret}, D., {Barcons}, X., {et~al.} 2013, ArXiv e-prints,
  arXiv:1306.2307

\bibitem[{{P{\'e}nin} {et~al.}(2012){P{\'e}nin}, {Lagache}, {Noriega-Crespo},
  {Grain}, {Miville-Desch{\^e}nes}, {Ponthieu}, {Martin}, {Blagrave}, \&
  {Lockman}}]{Penin:2012}
{P{\'e}nin}, A., {Lagache}, G., {Noriega-Crespo}, A., {et~al.} 2012, \aap, 543,
  A123

\bibitem[{{{\'S}liwa} {et~al.}(2001){{\'S}liwa}, {Soltan}, \&
  {Freyberg}}]{Sliwa:2001}
{{\'S}liwa}, W., {Soltan}, A.~M., \& {Freyberg}, M.~J. 2001, \aap, 380, 397

\bibitem[{{Spergel} {et~al.}(2015){Spergel}, {Gehrels}, {Baltay}, {Bennett},
  {Breckinridge}, {Donahue}, {Dressler}, {Gaudi}, {Greene}, {Guyon}, {Hirata},
  {Kalirai}, {Kasdin}, {Macintosh}, {Moos}, {Perlmutter}, {Postman},
  {Rauscher}, {Rhodes}, {Wang}, {Weinberg}, {Benford}, {Hudson}, {Jeong},
  {Mellier}, {Traub}, {Yamada}, {Capak}, {Colbert}, {Masters}, {Penny},
  {Savransky}, {Stern}, {Zimmerman}, {Barry}, {Bartusek}, {Carpenter}, {Cheng},
  {Content}, {Dekens}, {Demers}, {Grady}, {Jackson}, {Kuan}, {Kruk}, {Melton},
  {Nemati}, {Parvin}, {Poberezhskiy}, {Peddie}, {Ruffa}, {Wallace}, {Whipple},
  {Wollack}, \& {Zhao}}]{Spergel:2015}
{Spergel}, D., {Gehrels}, N., {Baltay}, C., {et~al.} 2015, ArXiv e-prints,
  arXiv:1503.03757

\bibitem[{{Wright}(1998)}]{Wright:1998}
{Wright}, E.~L. 1998, \apj, 496, 1

\bibitem[{{Yue} {et~al.}(2013){Yue}, {Ferrara}, {Salvaterra}, {Xu}, \&
  {Chen}}]{Yue:2013}
{Yue}, B., {Ferrara}, A., {Salvaterra}, R., {Xu}, Y., \& {Chen}, X. 2013,
  \mnras, 433, 1556

\end{thebibliography}

\end{document}